\documentclass[prl,twocolumn,showpacs,preprintnumbers,reprint,amsmath, amssymb,superscriptaddress]{revtex4-1} 

\usepackage{wasysym,helvet}
\usepackage{graphicx} 
\usepackage{dcolumn}  
\usepackage{bm}       
\usepackage{times,mathptmx}

\usepackage{color}

\usepackage{soul}

\graphicspath{{figuras/}}

\begin{document}
\newif\ifnocolor
\nocolortrue

\newcommand{\bea}{\begin{eqnarray}}
\newcommand{\eea}{ \end{eqnarray}}
\newcommand{\bit}{\begin{itemize}}
\newcommand{\eit}{ \end{itemize}}

\newcommand{\be}{\begin{equation}}
\newcommand{\ee}{\end{equation}}
\newcommand{\ra}{\rangle}
\newcommand{\la}{\langle}
\newcommand{\U}{\widetilde{U}}
\newcommand{\brac}[1]{\langle #1|}
\newcommand{\bra}[1]{\langle #1}
\newcommand{\ket}[1]{|#1\rangle}
\newcommand{\ktimes}{\rangle\! \langle}
\newcommand{\op}[2]{|#1\ktimes #2|}
\newcommand{\opoo}{\op{0}{0}}
\newcommand{\opoi}{\op{0}{1}}
\newcommand{\opio}{\op{1}{0}}
\newcommand{\opii}{\op{1}{1}}
\newcommand{\rhoq}{\rho_{\rm q}}
\newcommand{\rhoe}{\rho_{\rm env}}
\newcommand{\env}{{\rm env}}
\newcommand{\qubit}{{\rm qubit}}
\newcommand{\Mp}{{\cal M}_{\rm p}}
\newcommand{\Mm}{{\cal M}_{\rm m}}
\newcommand{\NPC}{\xi}
\newcommand{\ANPC}{\langle \xi\rangle}


\def\bracket#1#2{{\langle#1|#2\rangle}}
\def\inner#1#2{{\langle#1|#2\rangle}}
\def\expect#1{{\langle#1\rangle}}
\def\e{{\rm e}}
\def\proj{{\hat{\cal P} }}
\def\tr{{\rm Tr}}
\def\H{{\hat H}}
\def\Hdag{{\hat H}^\dagger}
\def\Lop{{\cal L}}
\def\Ehat{{\hat E}}
\def\Edag{{\hat E}^\dagger}
\def\Shat{\hat{S}}
\def\Sdag{{\hat S}^\dagger}
\def\Ahat{{\hat A}}
\def\Adag{{\hat A}^\dagger}
\def\U{{\hat U}}
\def\Udag{{\hat U}^\dagger}
\def\Zhat{{\hat Z}}
\def\Phat{{\hat P}}
\def\Op{{\hat O}}
\def\id{{\hat I}}
\def\x{{\hat x}}
\def\P{{\hat P}}	
\def\Px{\proj_x}
\def\Pr{\proj_{R}}
\def\Pl{\proj_{L}}
\def\ODR{f_{_{\rm DR}}(t)}
\def\ODRn{O_{_{\rm DR}}(n)}
\newcommand{\equa}[1]{Eq.~(\ref{#1})}
\newcommand{\comm}[1]{{\color{googB}[#1]}}
\newcommand{\comp}[1]{{\bf #1}}
\newcommand{\mcH}{\mathcal{H}}
\renewcommand{\e}{\text{env}}
\newcommand{\s}{{\rm sys}}
\newcommand{\eref}[1]{Eq.~(\ref{#1})}
\newcommand{\Eref}[1]{Eq.~(\ref{#1})}
\newcommand{\Sdec}{S_{\rm dec}} 		
\newcommand{\SD}{S_{\rm D}} 			
\newcommand{\lambdac}{\lambda_{\rm c}}	
\newcommand{\IQS}{\textsf{IQS} }
\newcommand{\nmax}{n_{\rm max}}

\def\co#1{{\color{red}\textst{#1}}}
\newcommand{\aug}[1]{#1}
\definecolor{verde}{rgb}{0.458,0.765,0.25}
 \definecolor{googB}{rgb}{0.285,0.539,0.949}
 \definecolor{fbkblue}{rgb}{0.2304, 0.3476,  0.5937}
\newcommand{\nach}[1]{{\color{verde} #1}}

\title{Comment on ``Exploring chaos in the Dicke model using ground-state fidelity and the Loschmidt echo''} 
\author{Ignacio Garc\'{\i}a-Mata}
\affiliation{Instituto de Investigaciones F\'isicas de Mar del Plata (IFIMAR, CONICET), 
Universidad Nacional de Mar del Plata, Mar del Plata, Argentina.}
\email{i.garcia-mata@conicet.gov.ar}
\affiliation{Consejo Nacional de Investigaciones Cient\'ificas y Tecnol\'ogicas (CONICET), Argentina}
\author{Augusto J. Roncaglia}
\affiliation{\mbox{Departamento de F\'{\i}sica ``J. J. Giambiagi" and IFIBA, 
             FCEyN, Universidad de Buenos Aires, 1428 Buenos Aires, Argentina}}                         
\author{Diego A. Wisniacki}
\affiliation{\mbox{Departamento de F\'{\i}sica ``J. J. Giambiagi" and IFIBA, 
             FCEyN, Universidad de Buenos Aires, 1428 Buenos Aires, Argentina}}
\date{\today}

\begin{abstract} 
In  [PRE \textbf{90}, 022920 (2014)] a study of the ground state fidelity of the Dicke model as a function of the coupling parameter is presented.  
Abrupt jumps of the fidelity in the superradiant phase are observed and are assumed to be related to the transition to chaos. We 
show that this conclusion results from a misinterpretation of the numerics. In fact, if the parity symmetry is taken into account, the unexpected 
jumps disappear.
\end{abstract}

\maketitle
 In \cite{Bhattacharya2014} as study of the ground state fidelity of the Dicke Model (DM) is presented. 
The ground state fidelity is defined as
\begin{equation}
F=|\bra{\psi(\lambda_0)}\ket{\psi(\lambda_0+\delta\lambda)}|^2,
\end{equation}
which is the overlap between ground states with slightly different interaction strength.
 The DM describes the interaction of the global spin of $N$ spin $1/2$ particles with an external field. The Hamiltonian that describes this system is 
 \begin{equation}
H=\omega_0 J_z+\omega a^\dagger a +\frac{\lambda}{\sqrt{2 j}}(a^\dagger+a)(J_++J_-).
\end{equation}
For parameter values $\omega=\omega_0=1$ there is superradiant quantum phase transition
at $\lambda_c=0.5$. At approximately the same value the level spacing distribution from Poisson to Wigner-Dyson which can identified with an integrability-chaos transition. \cite{EmaryBrandes2003}

\begin{figure}
\includegraphics[width=\linewidth]{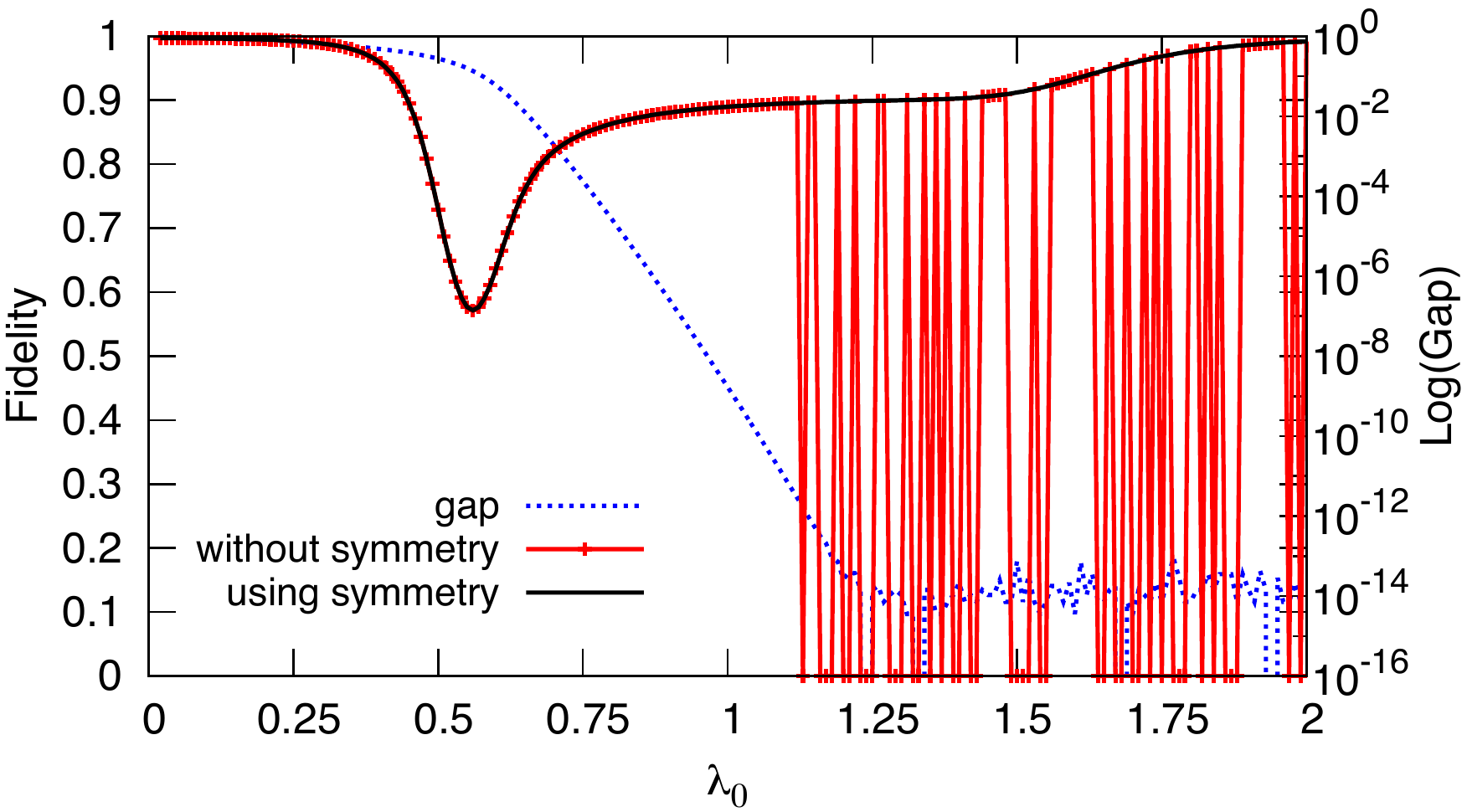}
\caption{Ground state fidelity for the DM with $J=5$ and $n_{\rm cut}=40$. Red line (and $+$ signs): without looking at parity of the ground state. Black solid line: taking into account parity flips. Dotted blue line (scale on the right y-axis): (log of the) gap $|E_0^{(even)}-E_0^{(odd)}|$ as a function of the coupling $\lambda_0$}
\end{figure}

The main numerical result shown in Fig. 3 of  
\cite{Bhattacharya2014}, abrupt jumps (so called ``aperiodic oscillations'') can be
observed and since they happen for values of $\lambda_0$ beyond the superradiant and integrability transition, the authors attribute these jumps 
to chaos. 
In what follows we provide a clear proof that this is not the case.

The DM has a well defined parity symmetry and the Hilbert space can be split into two noninteracting subspaces. 
When the ground state fielity is computed without taking into account the parity symmetry  
abrupt jumps can be observed in Fig. 1 (red line). These jumps are completely analogous to those presented in Fig. 3 of \cite{Bhattacharya2014}.
On the other hand, if we now take into account the parity symmetry, the jumps disappear completely (black curve of Fig. 1).

\begin{figure}
\includegraphics[width=\linewidth]{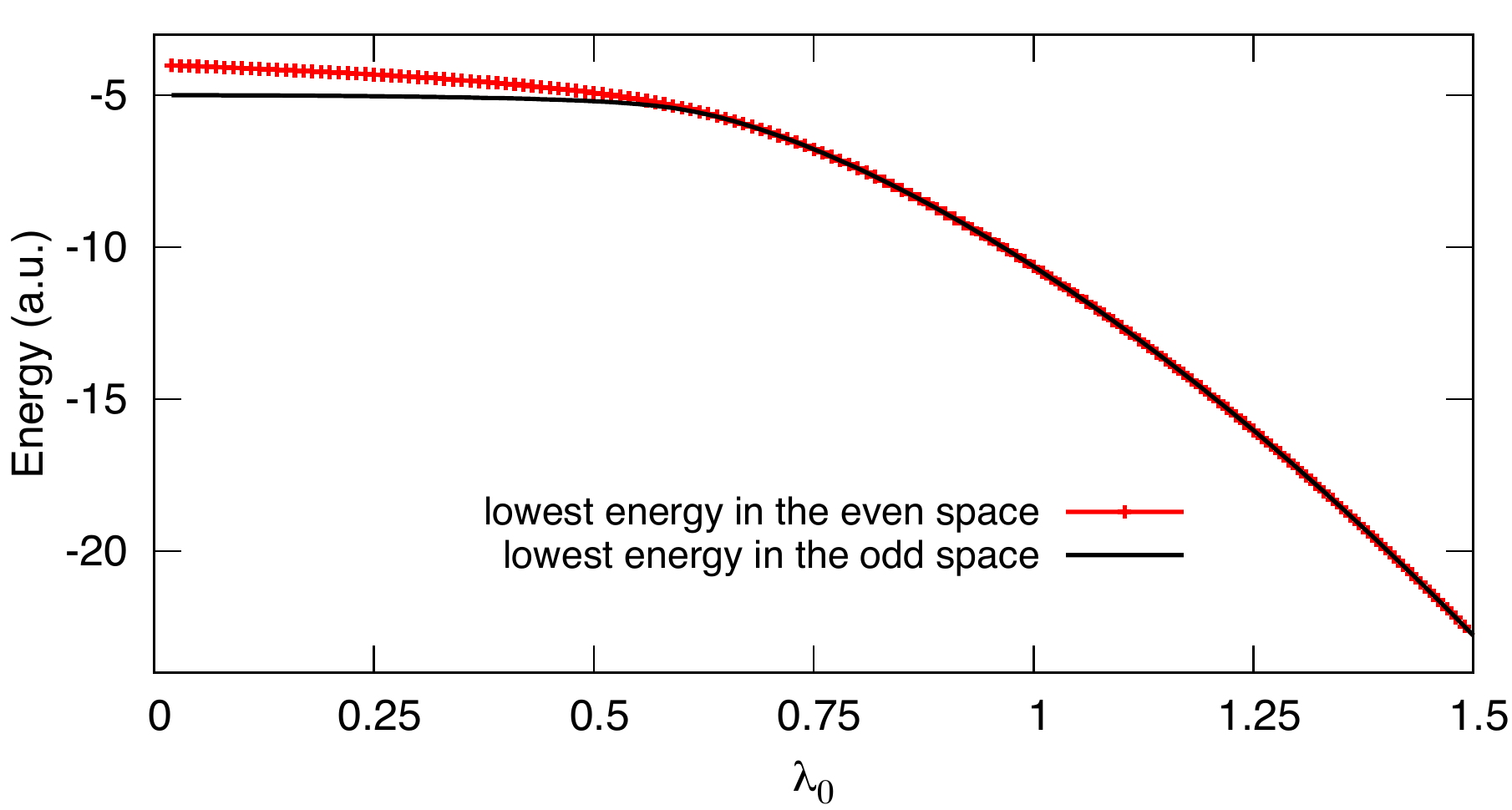}
\caption{Energy of the ground state, for both odd and even space, of the DM for $J=5$ and $n_{\rm cut}=40$. }
\end{figure} 
The curves presented in Fig. 1 clearly show that the jumps attributed to chaos in Ref. \cite{Bhattacharya2014}  are due to nothing but a numerical issue: for coupling values beyond the superradiant transition the system becomes gapless, i.e. the smallest energies corresponding to different parity states become machine-precision-degenerate (see Figs. 1 and 2).
Therefore, the diagonalization routine cannot resolve between them and from $\lambda_0$ to $\lambda_0+\delta\lambda$ it randomly yields ground states with different parities which of course are orthogonal (thus $F=0$). 
 In Fig. 2 we show the lowest energy for both even and odd spaces as a function of $\lambda_0$. It is clearly observed that after $\lambda_c$ both lower 
 energies, $E_0^{\rm (even)}$ and $E_0^{\rm (odd)}$, become increasingly close.
In Fig. 1 (blue curve) we plot the gap $|E_0^{\rm (even)}-E_0^{\rm (odd)}|$, and we see that it is precisely  when this difference reaches the order of machine precision ($\sim 10^{-14}$) is when the oscillations start to appear.

\noindent \textit{Acknowledgments.}
I.G.M. and D.A.W.
received support from ANPCyT (PICT 2010-1556), UBACyT,
and CONICET (PIP 114-20110100048 and PIP 11220080100728).
A.J.R. acknowledges support from CONICET and grants from ANPCyT 
(PICT 02843) and Ubacyt.

%

\end{document}